\documentclass[12pt]{article}
\usepackage{epsf}
\usepackage{pazh_en}
\usepackage{graphicx}
\tightenlines

\sloppypar

\begin{document}
\baselineskip 21pt


\title{\bf YOUNG NUCLEI IN DWARF ELLIPTICAL GALAXIES}

\author{\bf \hspace{-1.3cm}\copyright\, 2006 г. \ \ 
I.V.Chilingarian\affilmark{1*,3}, O.K.Sil'chenko\affilmark{1},
V.L.Afanasiev\affilmark{2}, Ph.Prugniel\affilmark{3}}

\affil{
{\it Sternberg Astronomical Institute of the Moscow State University,Universitetsky pr. 13, Moscow, 119992, Russia}$^1$\\ 
{\it Special Astrophysical Observatory of the Russian Academy of Sciences,
Nizhniy Arkhyz, Karachaevo-Cherkesia, 369167, Russia}$^2$\\
{\it Universit\'e de Lyon 1, Centre de Recherche Astronomique de Lyon, Observatoire de Lyon,
9 avenue Charles Andr\'e, F-69230 Saint-Genis Laval, France ; CNRS, UMR 5574 ; Ecole Normale Sup\'erieure de Lyon, Lyon, France}$^3$}

\vspace{2mm}
\received{10 November 2006}
\accepted{15 November 2006}

\sloppypar 
\vspace{2mm}
\noindent

We report discovery of young embedded structures in three diffuse elliptical
galaxies (dE) in the Virgo cluster: IC~783, IC~3468, and IC~3509. We performed 3D
spectroscopic observations of these galaxies using the MPFS spectrograph at
the Russian 6-m telescope, and obtained spatially resolved distributions of
kinematical and stellar population parameters by fitting high-resolution
PEGASE.HR synthetic single stellar populations (SSP) in the pixel space.
In all three galaxies, the luminosity weighted age of the nuclei, about 4~Gyr,
is considerably younger than population in the outer regions of the
galaxies. We discuss two possibilities to acquire the observed structures -- 
dissipative merger event and different ram pressure stripping efficiency
during two consequent crossings of the Virgo cluster centre.

\noindent
{\bf Keywords:\/}
galaxies: dwarf -- galaxies: evolution -- galaxies: elliptical and
lenticular, dE -- galaxies: stellar content -- galaxies: individual: IC~783, 
IC~3468, IC~3509.

\noindent
{\bf PACS codes:\/ 98.62.-g, 98.62.Ai, 98.62.Lv, 98.62.Bj, 98.52.Wz}

\vfill
\noindent\rule{8cm}{1pt}\\
{$^*$ e-mail: $<$chil@sai.msu.su$>$}

\clearpage

\section{INTRODUCTION}
\label{secintro}
Diffuse (or dwarf) elliptical galaxies represent a numerically dominated
population in dense regions of the Universe, but their origin and evolution still
remains unknown. They exhibit great variety of kinematical and stellar population
properties (Simien \& Prugniel 2002, Geha et al. 2002, 2003, De Rijcke et al. 2004,
Van Zee et al. 2004a,b), raising a number of questions to the present theories of 
galactic evolution. In the past, based on the photometrical analysis, dE
galaxies were believed to be old metal-poor objects, considered as building blocks
for larger galaxies, which was consistent with the hierarchical merger scenario.
However, spectroscopic studies (Geha et al. 2003, Van Zee et al. 2004b) pointed
out to mean luminosity-weighted ages of bright dE's in the Virgo cluster as
young as 3~Gyr, with a total absence of objects older than 10~Gyr.

Kinematics of dE's sometimes is very complex as well. Selected objects
exhibit kinematically decoupled cores (De Rijcke et al. 2004, Geha et al.
2005, Thomas et al. 2006).

Recently, high quality HST imagery became available for a large sample of 
early-type galaxies (dwarfs to giants) in the Virgo cluster
(C\^ot\'e et al. 2004). A population of compact blue nuclei was discovered in the
low- and moderate-luminosity galaxies (Ferrarese et al. 2006). Their g'-z'
colours were consistent either with younger ages, or with lower metallicities,
compared to their host galaxies.

Young nuclei are a frequent phenomenon in the giant early-type galaxies
(Sil'chenko 1997, Vlasyuk \& Silchenko 2000) observed both in clusters
and groups (Sil'chenko 2006). However, there were no detections of young
circumnuclear structures in dwarf elliptical/lenticular galaxies, probably
because of difficulties observing them due to low surface brightness.
To reach high signal-to-noise ratios, sufficient for the stellar population
analysis using classical approach by measuring Lick indices
(Worthey et al. 1994), integration time must be order of several hours with 
large telescopes. In addition, integral field spectroscopy is an essential 
technique for reliable detections of such structures. No attempts
have been made so far to observe even small samples of dE galaxies with IFU
spectrographs, but only individual objects (Geha et al. 2005).

We have started a project of observing a sample of dE galaxies in cluster
and groups using the Multi-Pupil Fiber Spectrograph at the Russian 6-m telescope BTA.

\section{Observations and data reduction}
\label{secobs}
The spectral data we analyse were obtained with the MPFS integral field
spectrograph. Idea of integral field spectroscopy was proposed by G.
Court\`es in the late 60th (for a description of the instrumental idea, see,
e.g., Bacon et al. 1995). It makes possible to obtain a set of spectra in a
wide spectral range of an extended area on the sky simultaneously.

The Multi-Pupil Fiber Spectrograph (MPFS), operated on the 6-m telescope
Bolshoi Teleskop Al'tazimutal'nij (BTA) of the Special Astrophysical
Observatory of the Russian Academy of Sciences, is a fibre-lens spectrograph
with a microlens raster containing $16 \times 16$ square spatial elements,
each with the size of 1''$\times$1'' (Afanasiev et al. 2001). We used
the setup of the instrument
providing intermediate spectral resolution, varying from $R=1300$ to
$R=2200$ over the field of view and the selected spectral range (4100 --
5650\AA). Parameters of the observations are summarized in Tab\~ref{tabobsdEV}.

\begin{table}
\begin{tabular}{lccc}
Object & Dates & Total exp.time & Seeing \\
\hline
IC~783 & 21,23 May 2004 & 3.5h & 2'' \\
IC~3468 & 20 March 2004 & 2.5h & 1.5'' \\
IC~3509 & 10,12 May 2005 & 3.5h & 1.7'' \\
\hline
\end{tabular}
\caption{Parameters of observations \label{tabobsdEV}}
\end{table}

The data reduction for 3D spectroscopy is a very elaborated process. We
are referring to our paper (Chilingarian et al., 2006 submitted) for a
detailed description of this procedure. The result of the data reduction
is flux calibrated spectra and parametrized line spread function (LSF) of the
spectrograph for every spatial element at every wavelength.

We apply Voronoi 2D adaptive binning procedure (Cappellari \& Copin, 2003)
in order to achieve sufficient signal to noise ratio of the data for extracting
stellar kinematics and stellar population parameters ($S/N=15\ldots20$) by
degrading spatial resolution. The adaptive binning procedure results in a
set of 1D spectra, for those all further steps of the analysis might be done
independently.

Besides we use an alternative tessellation of the datasets containing two
bins (''2-points'' hereafter): central young embedded structure, and the
rest of the galaxy.

For two galaxies: IC~3468 and IC~3509 we have used results of analysis of
ACS images from the HST archive, proposal 9401, "The ACS Virgo Cluster Survey"\ 
(P.I.: P. C\^ot\'e), presented in Ferrarese et al. (2006). For IC~783 we have
used light and colour profiles available through the GOLDMine database
(Gavazzi et al. 2003).

To estimate stellar population parameters we exploit the new innovative
approach developed in our team. Its idea is the simultaneous determination
of the stellar population and kinematical parameters by pixel space fitting
of observed spectra of galaxies in every spatial bin by high-resolution
PEGASE.HR synthetic spectra (Le Borgne et al. 2004) broadened corresponding
to the line-spread-function (LSF) of the spectrograph. Details of the method
are given in Chilingarian et al. (2005), Prugniel et al. (2005), its
stability and biases specifically for MPFS observations are discussed in
details in Chilingarian et al. (2006 submitted). Summary of derived 
kinematical and stellar population parameters of thee galaxies is given in
Table~\ref{tabparams}.

\section{Stellar population and internal kinematics}

\begin{table}
\begin{tabular}{lccc}
Object & $t$ Gyr & $Z$ dex & $(M/L)_{B*}$ \\
\hline
IC~783 (core) & $3.3\pm0.4$ & $-0.35\pm0.04$ & $2.1\pm0.2$ \\
IC~783 (out.) & $12.8\pm4.0$ & $-0.79\pm0.12$ & $5.2 \pm 1.5$ \\
\hline
IC~3468 (disc) & $5.3\pm0.4$ & $-0.40\pm0.05$ & $2.8\pm0.4$ \\
IC~3468 (out.) & $8.6\pm0.9$ & $-0.60\pm0.05$ & $4.0\pm0.6$ \\
\hline
IC~3509 (core) & $4.1\pm0.4$ & $-0.05\pm0.05$ & $3.1\pm0.4$ \\
IC~3509 (out.) & $7.8\pm0.8$ & $-0.40\pm0.10$ & $4.3\pm0.5$ \\
\hline
\end{tabular}
\caption{Luminosity-weighted parameters of the stellar populations of
three dE galaxies: age, metallicity [Fe/H], and mass-to-light ratios
of the stellar population according to Worthey (1994).
\label{tabparams}}
\end{table}
     
\subsection{IC~783}
\begin{figure}
\hfil
\begin{tabular}{c c}
 (a) & (b) \\
 \includegraphics[width=7cm]{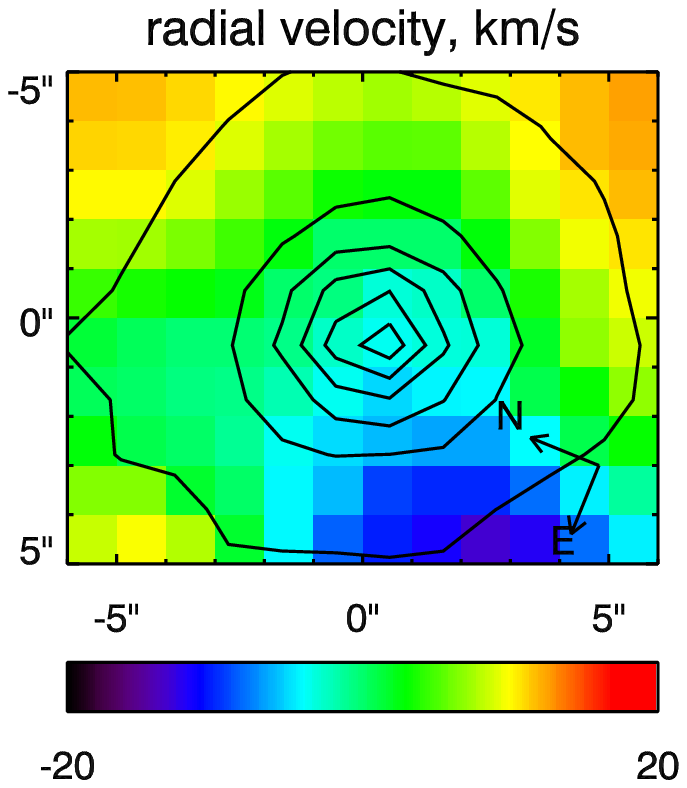} &
 \includegraphics[width=7cm]{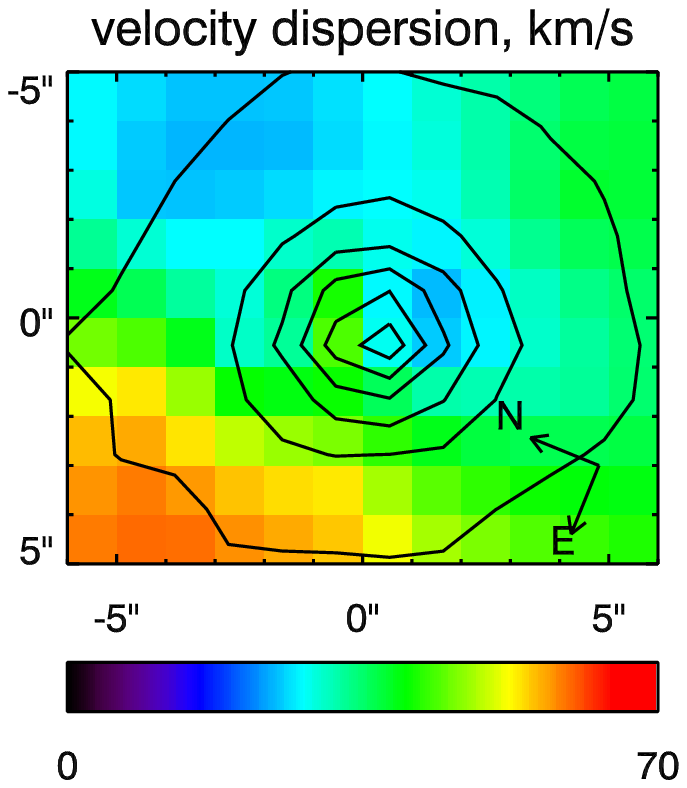} \\
 (c) & (d) \\
 \includegraphics[width=7cm]{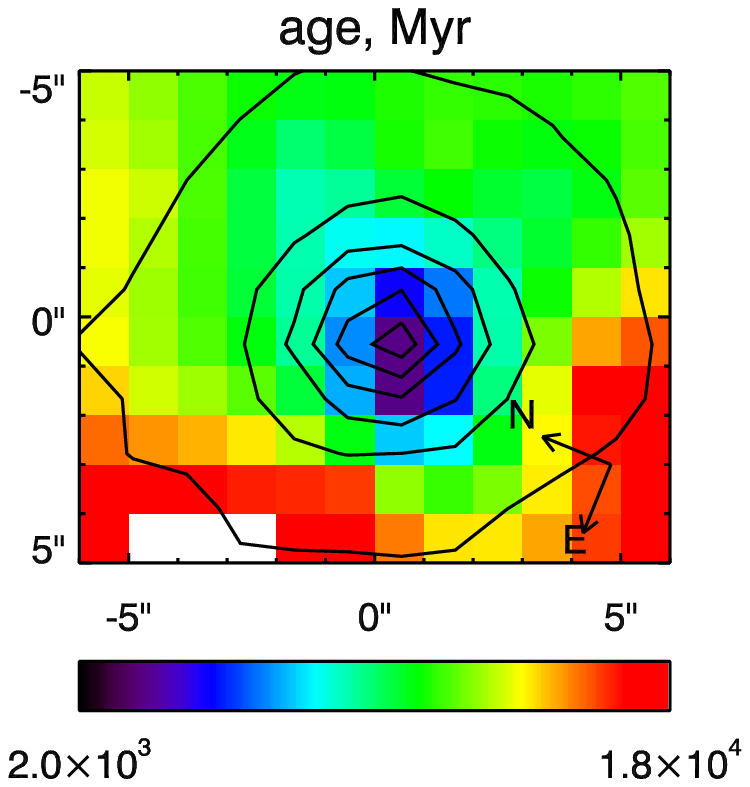} &
 \includegraphics[width=7cm]{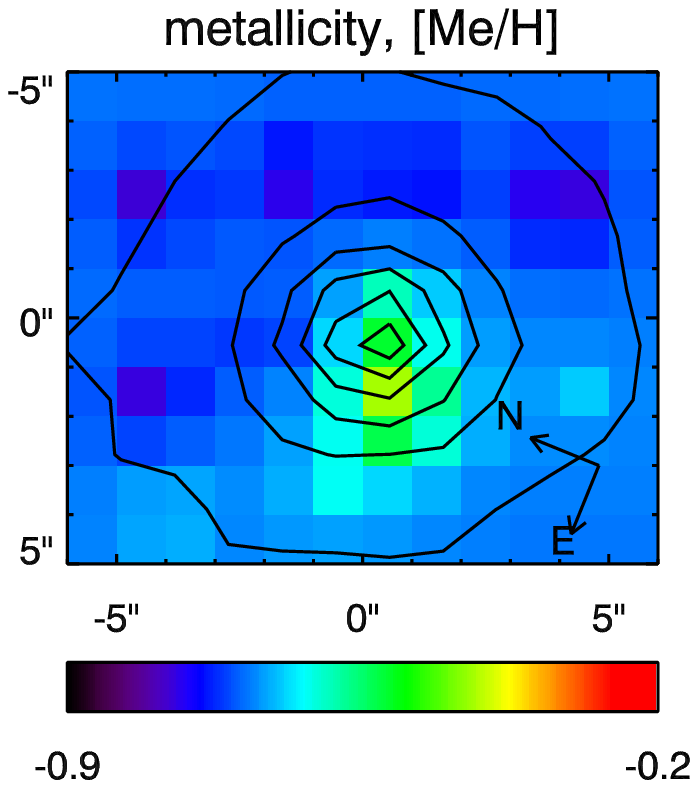} \\
\end{tabular}
\caption{Kinematics and stellar population of IC~783. Maps of internal kinematics
and stellar population parameters are built for a Voronoi tessellation with
a target S/N ratio or 15. (a) line-of-sight stellar velocity, (b) stellar
velocity dispersion, (c) luminosity-weighted age, (d) luminosity-weighted
metallicity.\label{figic783}}
\end{figure}

This galaxy was found to have a remarkable spiral structure (Barazza et al.
2002). IC~783 exhibits rotation ($v_{rot} \sim 20$~km~s$^{-1}$, see also
Simien \& Prugniel, 2002). Velocity dispersion field is flat
($\langle \sigma \rangle =35\pm 10$~km~s$^{-1}$)
and does not show any significant features (Fig.~\ref{figic783}).

We have discovered a young nucleus in this galaxy, having luminosity-weighted
stellar population parameters: $t=3.3\pm0.4$~Gyr,
$Z=-0.35\pm0.04$~dex -- compared to the main body
(containing the spiral structure): $t=12.8\pm4.0$~Gyr, $Z=-0.79\pm0.12$~dex. 
Young nucleus in IC~783 remains spatially
unresolved. The B-band mass-to-light ratio for the stellar populations
according to Worthey (1994) are: $(M/L)_{B*} = 2.1 \pm 0.2$ for the core,
and $(M/L)_{B*} = 5.2 \pm 1.5$ for the rest of IC~783.

\subsection{IC~3468}
\begin{figure}
\hfil
\begin{tabular}{c c}
 (a) & (b) \\
 \includegraphics[width=7cm]{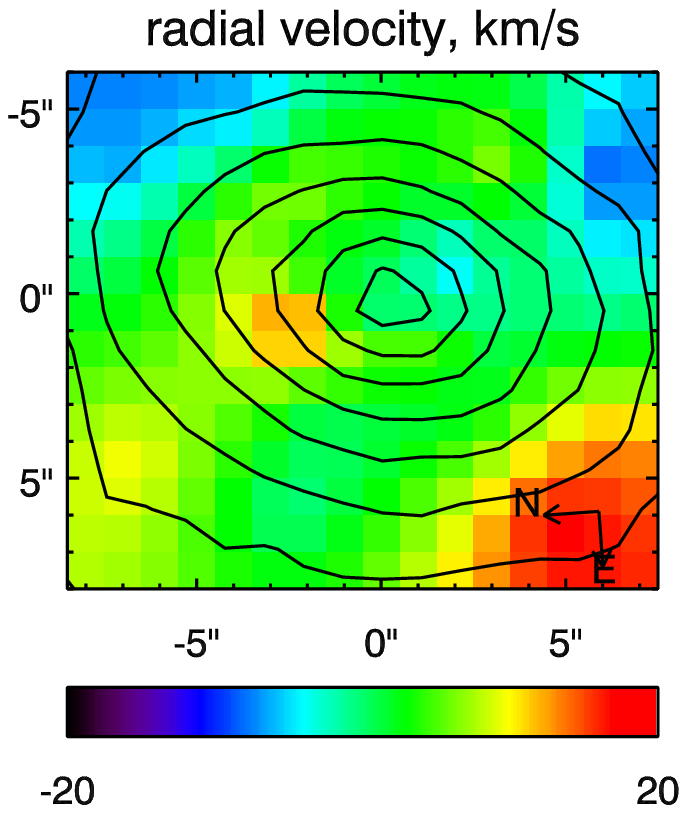} &
 \includegraphics[width=7cm]{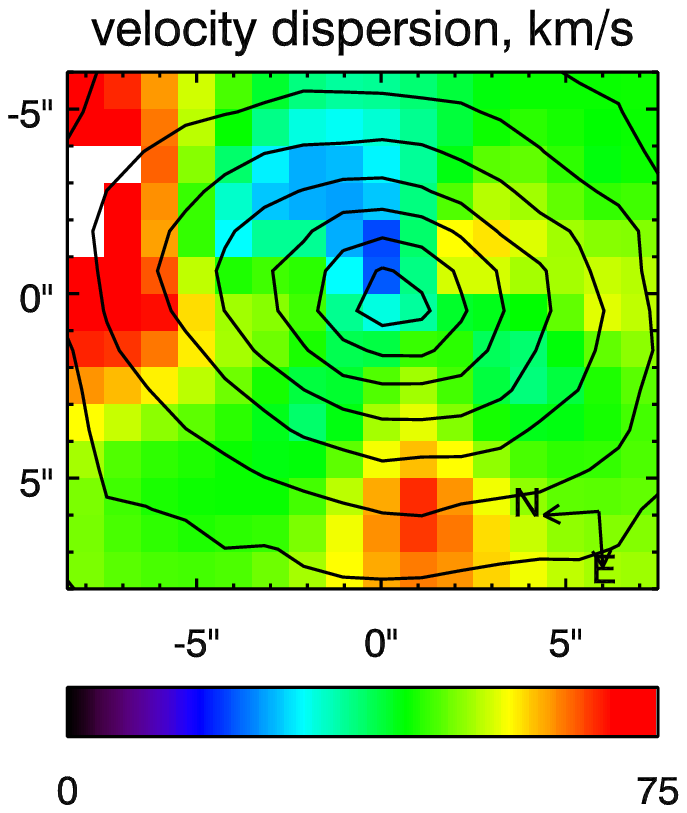} \\
 (c) & (d) \\
 \includegraphics[width=7cm]{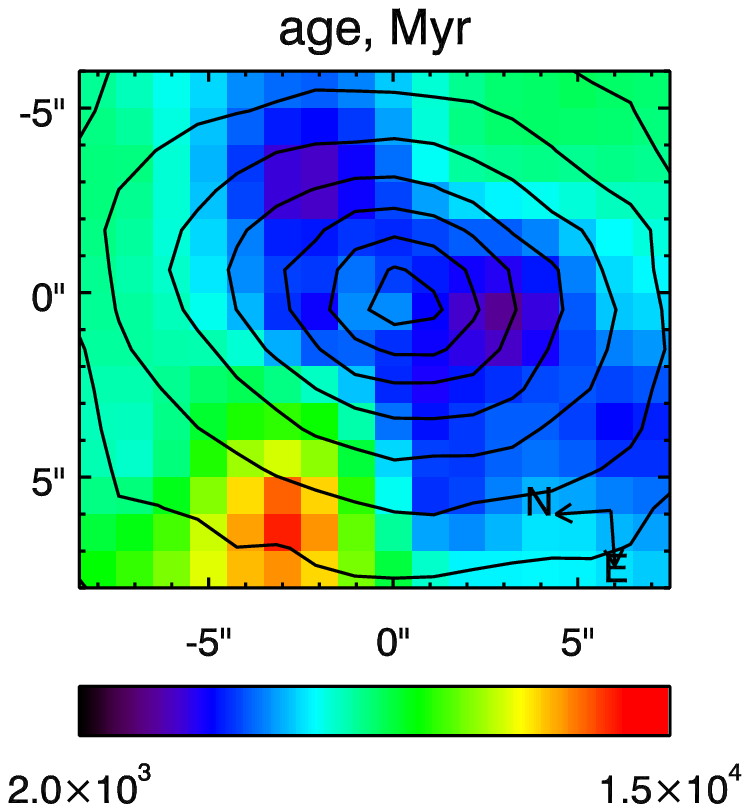} &
 \includegraphics[width=7cm]{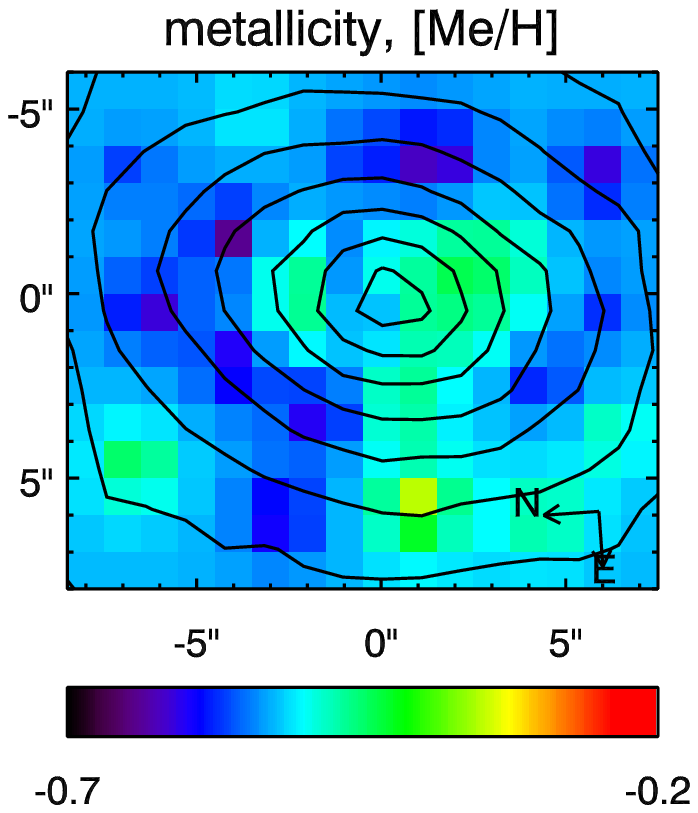} \\
\end{tabular}
\caption{
Kinematics and stellar population of IC~3468. Maps of internal kinematics
and stellar population parameters are built for a Voronoi tessellation with a
target S/N ratio or 15. (a) line-of-sight stellar velocity, (b) stellar velocity
dispersion, (c) luminosity-weighted age, (d) luminosity-weighted metallicity.
\label{figic3468}}
\end{figure}

An embedded structure is known to present in IC~3468 (Barazza et al. 2002).
The question on the nature of this substructure was left open, because no
rotation was detected in long-slit spectroscopic data by Simien \& Prugniel
(2002). However, we found a complex kinematics (Fig.~\ref{figic3468}) --
rotation along two non-perpendicular directions (NW-SE, and NE-SW separated
by $\sim$ 60 degrees), none of them coinciding with the position of the slit
from Simien \& Prugniel (2002). Unsharp masking of HST ACS imagery reveals
elongated structure in the central region of the galaxy. By varying the
smoothing radius for unsharp masking, different parts of this structure can
be revealed.

In the map of a luminosity-weighted stellar age, an elongated substructure having
$t=5.3\pm0.4$~Gy, coinciding with one of the rotating components (NW-SE) is
clearly seen. It is roughly 3.3~Gy younger than the rest of the galaxy
($t=8.6\pm0.9$~Gy). One may also notice a "blue stripe"\ in the velocity
dispersion distribution (lower by $\sim$ 10~km~s$^{-1}$ compared to mean
values) having the same locus. Based on these results we conclude that
NW-SE rotation corresponds to a moderately inclined stellar disc
($i \sim 60^{\circ}$). Rotational velocity is $17\pm4$~km~s$^{-1}$ at 7~arcsec
from the centre, but we cannot be sure to reach the maxima of rotation --
wider-field observations are needed.

Surprisingly, this disc almost does not affect the metallicity distribution.
Luminosity-weighted metallicity exhibits relatively smooth map
($Z=-0.60\pm0.05$~dex) with a slight gradient towards the centre
(up to $Z=-0.40\pm0.05$~dex).

The B-band mass-to-light ratios of the stellar populations are:
$(M/L)_{B*} = 2.8 \pm 0.4$ for the disc and $(M/L)_{B*} = 4.0 \pm 0.6$ for
outskirts. Our estimates of the luminosity-weighted age and metallicity in the
centre of IC~3468 are consistent with the g'-z' colour of the compact nucleus
(averaged within the host galaxy over the aperture corresponding to the seeing
conditions) provided in Ferrarese et al. (2006)

\subsection{IC~3509}
\begin{figure}
\hfil
\begin{tabular}{c c}
 (a) & (b) \\
 \includegraphics[width=7cm]{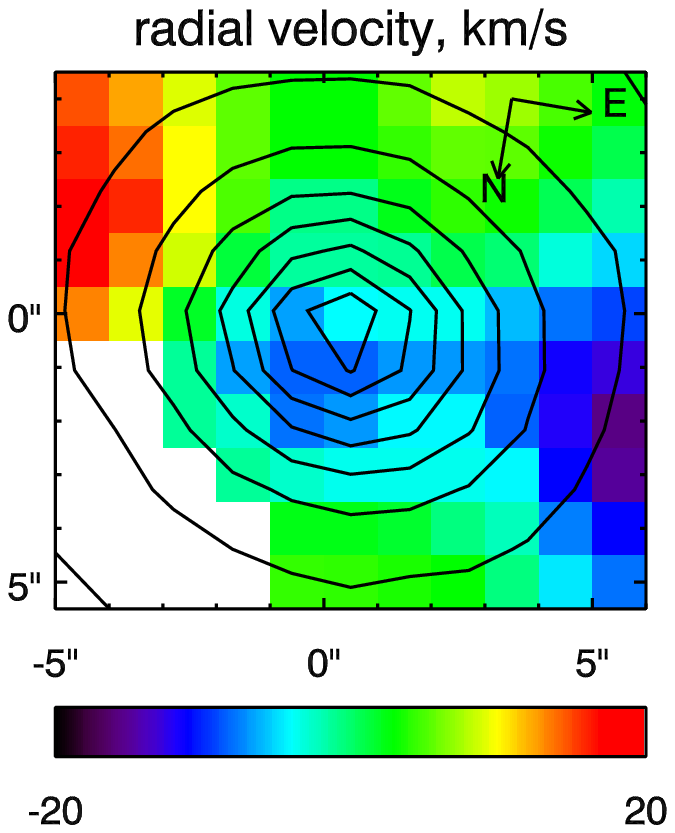} &
 \includegraphics[width=7cm]{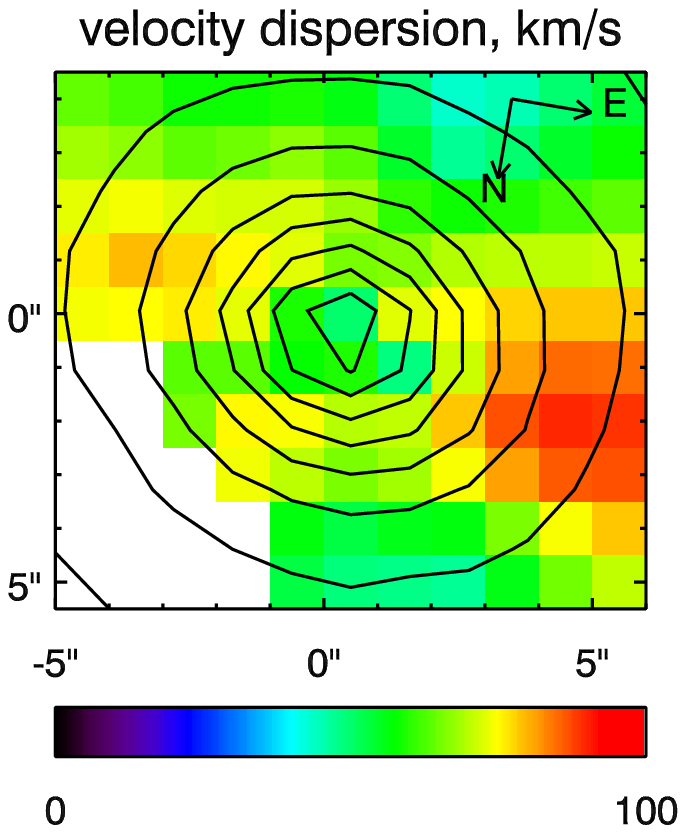} \\
 (c) & (d) \\
 \includegraphics[width=7cm]{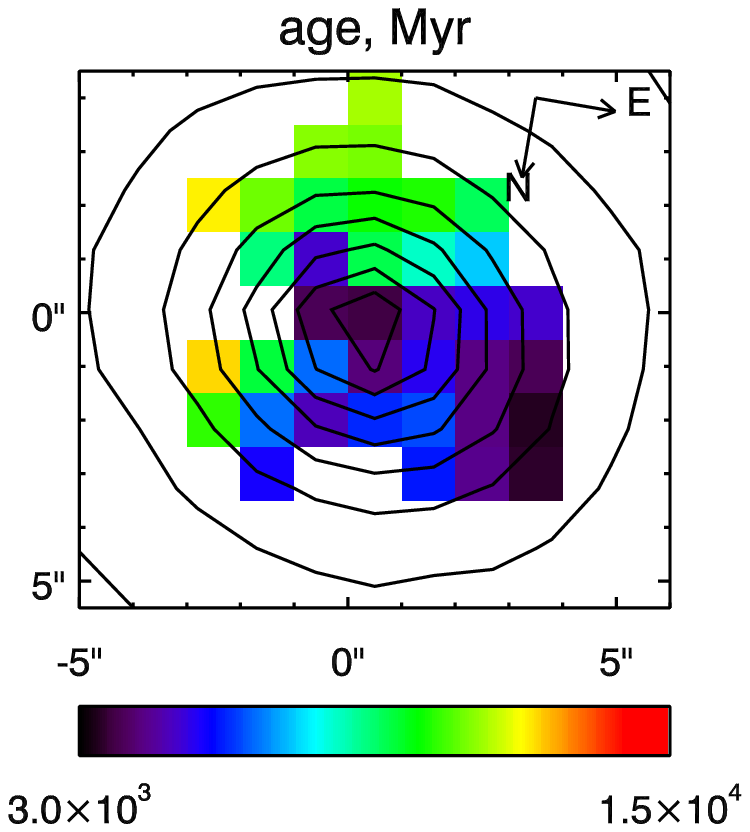} &
 \includegraphics[width=7cm]{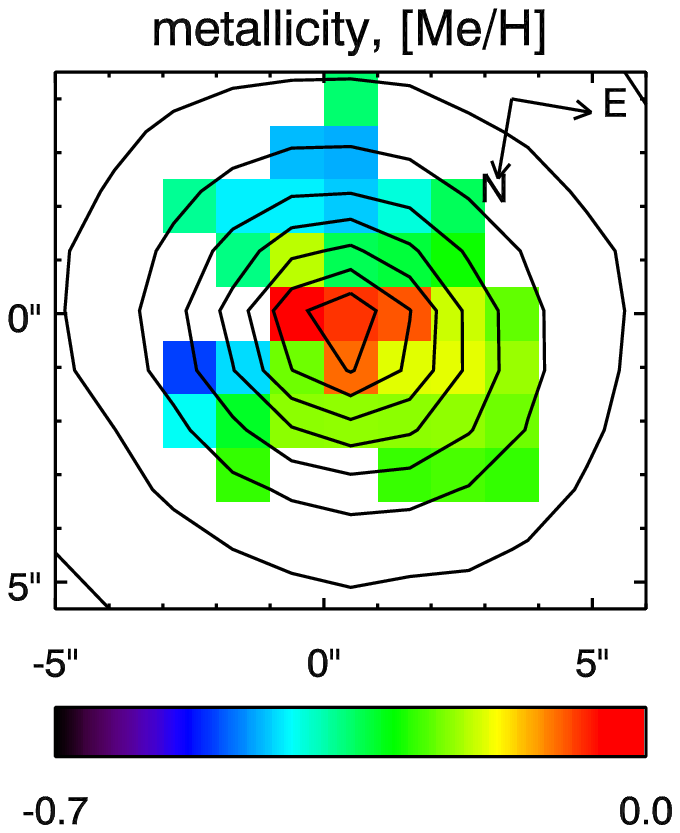} \\
\end{tabular}
\caption{
Kinematics and stellar population of IC~3509. Maps of internal kinematics
and stellar population parameters are built for a Voronoi tessellation with a
target S/N ratio or 10. (a) line-of-sight stellar velocity,
(b) stellar velocity dispersion, (c)
luminosity-weighted age, (d) luminosity-weighted metallicity.
\label{figic3509}}
\end{figure}

When we were selecting the targets for observations, IC~3509 was chosen as a
"prototypical"\ dE galaxy, classified as a galaxy without nucleus in
Binggeli et al. (1985). We did not expect to find unusual kinematics and/or
stellar populations in this object. However, we detected a kinematically
decoupled central region, rotating ($v_{rot} \sim 10$~km~s$^{-1}$) in the
perpendicular direction to the major axis (Fig.~\ref{figic3509}), where
significant rotation along the photometric major axis is also seen ($v_{rot}
\sim 20$~km~s$^{-1}$). This structure is associated with a dip in the
velocity dispersion distribution (50~km~s$^{-1}$ compared to 75~km~s$^{-1}$
at 4~arcsec from the centre) and a metallicity gradient of about 0.2~dex.
Stellar population of the galaxy is relatively old and metal-poor
($t=7.8\pm0.8$~Gyr, $Z=-0.40\pm0.10$~dex, $(M/L)_{B*} = 4.3 \pm 0.5$) In the
very centre of the galaxy we see a spatially unresolved young
($t=4.1\pm0.4$~Gyr) metallic ($Z=-0.05\pm0.05$~dex) nucleus ($(M/L)_{B*} =
3.1 \pm 0.4$).

We applied unsharp masking technique to the HST imagery available from the
Virgo ACS Survey (Cot\^e et al. 2004) with different smoothing radii.
No fine structures have been revealed.

Kinematical appearance quite similar to IC~3509 was observed earlier in giant
early-type galaxies, for example, in NGC~5982 (Statler 1991). An explanation
was proposed, which did not require presence of dynamically distinct structures --
projection of orbits in the triaxial potential. Based on quite regular (except
the very centre) maps of stellar population parameters of IC~3509 we
conclude that the galaxy outside the core region can be represented by a
single-component triaxial ellipsoid. 

As in the case of IC~3468, for IC~3509 nucleus, the g'-z' colour reported by
Ferrarese et al. (2006) is consistent with our stellar population parameter
estimations.

\section{Discussion}
\label{secdisc}
Since first discoveries of chemically (Sil'chenko et al. 1992) and
evolutionary (Sil'chenko 1997, Vlasyuk \& Silchenko 2000) decoupled cores in
giant early-type galaxies, no attempts of quantitative modelling of their
formation and evolution were made. Usual qualitative explanation of this
phenomenon is a dissipative merger event. Whereas merger is an established
scenario of formation for giant early-type galaxies, normally it is
considered as improbable for dwarfs because of their little sizes and masses.

However, presence of embedded disc in IC~3468 is an important argument for a
merger scenario. Kinematically decoupled structures, associated with young
metal-rich stellar population are consistent with a hypothesis of
dissipative merger event which took place several Gyr ago. Such a
dissipative merger is expected to trigger a starburst that consumes
available gas, leading to a kinematically decoupled core presented by the
circumnuclear stellar disc that is younger than the host galaxy.

Another possibility is effects of ram pressure stripping by the
intergalactic medium. Efficiency of ram pressure stripping depends on the
density of the region being stripped: higher density results in lower
efficiency (Gunn \& Gott, 1972; Abadi et al. 1999). Thus, it might happen
that in the dense nucleus of a dwarf galaxy gas will not be removed. A
similar phenomenon of gaseous disc truncation is observed in giant spiral
galaxies in the Virgo cluster (Cayatte et al. 1994, Kenney \& Koopmann 1999)
and modelled by Abadi et al. (1999). Gas, survived in the central region and
additionally compressed by the ram pressure stripping may rapidly form stars
and be transformed into a circumnuclear stellar disc.

A possible scenario to acquire a structure observed in IC~783 (full absence
of gas and young nucleus without obvious evidences of kinematical
decoupling) is multiple crossings of the cluster centre. IC~783 is located
on the projected distance of 1.1~Mpc from the centre of the Virgo cluster.
Thus its orbital period is at least 4.5~Gyr (assuming mass of the cluster of
$10^{14}$~M$_{\odot}$). Gas of IC~783 can be depleted in the disc during the
first passage, but preserved in the inner dense nucleus, because
intercluster medium density and/or velocity of the galaxy might be not
sufficient to remove gas completely.  To remove gas and stop star formation
in the core we may assume that during the second crossing several Gyr later
the intracluster orbit of the galaxy might be transformed into more
elongated one, say due to casual encounter with a massive galaxy, so
$v_{cross}$ would increase (and $\rho$ as well, because the galaxy would
pass closer to the centre of the cluster), resulting in ram pressure
$P = \rho v^2$ reaching sufficient value to strip the nuclear region of
the galaxy and stop star formation.

Alternative possibility of varying efficiency of ram pressure stripping
for IC~783 might be explained by its belonging to Messier~100
group. IC~783 is located at some 90~kpc of projected distance from M~100,
and radial velocity difference of $\sim$270~km~s$^{-1}$ is an argument for
Assuming that M~100 group is passing
near its apocentre now its orbital period in the cluster
turns to be about 5--8~Gyr, thus the recent crossing of the cluster centre
took place some 3--4~Gyr ago, and the previous one around 8--12~Gyr ago.
On the other hand, the orbital period of IC~783 with respect to
M~100 should be around 1~Gyr. Thus if the orbital velocity of IC~783 was
counter-directed to the orbital motion of M~100 in the cluster during the
first crossing of the central region, and co-directed during the second
crossing, the ram pressure value $P = \rho v^2$ might differ by a factor
of 3.5 (assuming the maximum velocity of M~100 with respect to
the Virgo intracluster medium to be $\sim 1000$~km~s$^{-1}$ and the orbital
velocity of IC~783 with respect to M~100 to be $\sim 300$~km~s$^{-1}$).
The coincidence of the estimated dates of the cluster centre crossing
by the M~100 group with the ages of two subpopulations in IC~783
is a strong argument for this scenario.

Ram pressure stripping during repetitive crossings of the cluster centre may
be considered as a possible explanation of young metal-rich cores in the
low-luminosity early type galaxies. Depending on the orbital parameters for
a particular galaxy, one would expect large scatter of ages/metallicities of
these substructures with respect to their host galaxies. This effect is able
explain increase of the scatter of average ages and metallicities of
early-type galaxies toward low-massive objects, as reported by Caldwell et
al. (2003) based on the results of multi-object spectroscopy.

Good agreement between our estimations of the stellar population parameters
and g'-z' colours in the nuclei of IC~3468 and IC~3509 may be considered as
an evidence for the presence of young metal-rich stellar populations in all
compact blue nuclei of dE galaxies (C\^ot\'e et al. 2006). However, this
hypothesis may only be proved by the forthcoming observations.

We thank A. V. Moiseev for supporting the observations with
the Multi-Pupil Fiber Spectrograph of the 6-m telescope.
Visits of PP in Russia and IC in France were supported through a CNRS
grant. PhD of IC is supported by the INTAS Young Scientist Fellowship
(04-83-3618). Special thanks to the Large Telescopes Time Allocation
Committee or the Russian Academy of Sciences for providing observing time
with MPFS. The 6-m telescope is
operated under the financial support of the Science Ministry of Russia
(registration number 01-43). During our data analysis
we used the Lyon-Meudon Extragalactic Database (LEDA) supplied by the
LEDA team at the CRAL-Observatoire de Lyon (France) and the NASA/IPAC
Extragalactic Database (NED) operated by the Jet Propulsion
Laboratory, California Institute of Technology under contract with
the National Aeronautics and Space Administration. The work
is partially based on the observations made with the NASA/ESA
Hubble Space Telescope, obtained from the data archive at the Space
Telescope Science Institute. STScI is operated by the Association of
Universities for Research in Astronomy, Inc. under the NASA contract
NAS 5-26555. The work on the study of dwarf galaxies in Virgo
is supported by the bilateral grant RFBR-Flanders 05-02-19805-MF\_a.

\pagebreak   

\newpage

\newpage

\newpage

\end{document}